\documentclass[conference,10pt]{IEEEtran}
\IEEEoverridecommandlockouts

\usepackage[numbers,sort&compress]{natbib}
\def\BibTeX{{\rm B\kern-.05em{\sc i\kern-.025em b}\kern-.08em T\kern-.1667em\lower.7ex\hbox{E}\kern-.125emX}}

\usepackage{amsmath,amssymb,amsfonts}

\usepackage{graphicx}
\usepackage{booktabs}
\usepackage{multirow}
\usepackage{adjustbox}
\usepackage{stfloats}
\usepackage{enumitem}

\usepackage[ruled,vlined,linesnumbered]{algorithm2e}

\usepackage{hyperref}
\usepackage{url}

\usepackage{textcomp}
\usepackage{nicefrac}
\usepackage{microtype}
\usepackage{xcolor}

\usepackage{tikz}

\usepackage{orcidlink}
\usepackage{comment}
\usepackage{braket}
\usepackage{caption}
\usepackage{multicol}
\usepackage{multirow}

\begin{document}

\bstctlcite{IEEEexample:BSTcontrol} 

\title{
Quantum Bayesian Networks for Machine Learning in Oil-Spill Detection
\vspace{10pt}
}

\vspace{10pt}
\author{\IEEEauthorblockN{\textbf{Owais Ishtiaq Siddiqui}\textsuperscript{1}, \textbf{Nouhaila Innan}\textsuperscript{2,3}, \textbf{Alberto Marchisio}\textsuperscript{2,3},\\ \textbf{Mohamed Bennai}\textsuperscript{4}, \textbf{Muhammad Shafique}\textsuperscript{2,3}
} \IEEEauthorblockA{
\textsuperscript{1}Department of Physics, COMSATS University Islamabad, Pakistan\\ 
\textsuperscript{2}eBRAIN Lab, Division of Engineering, New York University Abu Dhabi (NYUAD), Abu Dhabi, UAE\\ \textsuperscript{3}Center for Quantum and Topological Systems (CQTS), NYUAD Research Institute, NYUAD, Abu Dhabi, UAE\\ 
\textsuperscript{4}Quantum Physics and Spintronics Team, LPMC, Faculty of Sciences Ben M'sick,\\ Hassan II University of Casablanca, Morocco\\ fa21-rph-029@isbstudent.comsats.edu.pk, nouhaila.innan@nyu.edu, alberto.marchisio@nyu.edu,\\ mohamed.bennai@univh2c.ma, muhammad.shafique@nyu.edu\\ }}

\maketitle

\begin{abstract}

Quantum Machine Learning (QML) has shown promise in diverse applications such as environmental monitoring, healthcare diagnostics, and financial modeling. However, its practical implementation faces challenges, including limited quantum hardware and the complexity of integrating quantum algorithms with classical systems. One critical challenge is handling imbalanced datasets, where rare events are often misclassified due to skewed data distributions. Quantum Bayesian Networks (QBNs) address this issue by enhancing feature extraction and improving the classification of rare events such as oil spills. This paper introduces a Bayesian approach utilizing QBNs to classify satellite-derived imbalanced datasets, distinguishing ``oil-spill'' from ``non-spill'' regions. QBNs leverage probabilistic reasoning and quantum state preparation to integrate quantum enhancements into classical machine learning architectures. Our approach achieves a 0.99 AUC score, demonstrating its efficacy in anomaly detection and advancing precise environmental monitoring and management. While integration enhances classification performance, dataset-specific challenges require further optimization.
\end{abstract}


\section{Introduction}
Effective environmental monitoring is paramount for safeguarding natural resources and ensuring prompt responses to ecological disasters. Oil-spills are one of the most significant environmental threats, with profound impacts on marine ecosystems and the economic health of coastal communities. 
The ability to detect these spills both accurately and with minimal latency is crucial for the early deployment of countermeasures and for mitigating long-term ecological damage that may persist for decades.

Recent events underscore the ongoing challenge and high stakes associated with oil-spills. For instance, in February 2023, the MT Princess Empress, carrying 900,000 liters of industrial oil, capsized off the coast of Naujan, Oriental Mindoro
\cite{crecy2024mt}. This major spill posed severe risks to the local marine environment and threatened the livelihoods of surrounding coastal communities due to potential impacts on tourism and fishing industries. Looking back, the 2010 Deepwater Horizon oil-spill, which released approximately 210 million gallons of oil into the Gulf of Mexico \cite{nixon2016shoreline}, and the 2020 Arctic Lake Pyasino spill, involving 20,000 tons of diesel \cite{hettithanthri2024review}, highlight that significant spills are a global and persistent issue.

Detecting oil-spills when they occur, presents numerous challenges. Oil can spread rapidly over large water areas, often influenced by currents and weather conditions, which complicates tracking and containment efforts. The varying types and thicknesses of oil further affect the visibility and spectral signatures detectable by remote sensing technologies. Moreover, the remote locations of many spills, like the deep-sea site of the Deepwater Horizon disaster, can severely delay detection and response efforts, exacerbating environmental impacts. Therefore, advanced detection methods that can swiftly identify and assess the scope of spills are essential for enabling more effective response strategies and ultimately reducing the ecological and economic damage caused by these disasters.

Recent advancements in Quantum Machine Learning (QML) provide new opportunities to enhance environmental monitoring by processing complex, large-scale data more effectively \cite{biamonte2017quantum, zaman2023survey, rezaei2024environmental, marchisio2024cutting, kashif2022demonstrating,kashif2024computational}. QML merges Quantum Computing (QC) principles with classical Machine Learning (ML) algorithms, offering potential advantages in speed and handling of high-dimensional data \cite{yamasaki2020learning,caro2022generalization, Innan_Grover_2024, innan2023enhancing, farhi2018classification, maouaki2024designing, innan2024financial, innan2024fedqnn, el2024quantum, dutta2024qadqn,innan2024qfnn,innan2024fl,dutta2024mqfl,maouaki2025qfal,dutta2024federated,chen2024crossing,innan2024quantum1,innan2025qnn,innan2025optimizing}. Despite challenges such as limited quantum hardware availability, noise in quantum systems \cite{CerezoNature2022, HuangHS2022}, and difficulties in merging quantum algorithms with existing classical systems \cite{peral2024systematic,liu2024laziness, el2024robqunns, ahmed2025quantum, el2024advqunn}, the unique properties of QML make it particularly well-suited for analyzing satellite-derived data for oil-spill detection \cite{Alam2021}. 

Another significant challenge in environmental monitoring is the imbalance in datasets \cite{tang2024review, LITTLE2021105509}, where instances of anomalies like oil-spills are relatively rare compared to normal conditions. This imbalance often results in biased models that struggle to detect anomalies accurately. Bayesian methods, known for their strength in probabilistic reasoning and uncertainty management, can help address this issue. Incorporating Bayesian approaches within QML, specifically through Quantum Bayesian Networks (QBNs), offers potential advantages, including enhanced computational power, accuracy, and efficient resource utilization \cite{SimaEQCE2020}. QBNs have demonstrated exceptional performance across various domains \cite{wang2024quantum,borujeni2021quantum,harikrishnakumar2023forecasting,carrascal2023bayesian}, motivating our exploration of their potential in environmental applications. These benefits contribute to improved classification performance on imbalanced datasets, making QBNs a preferred tool for advanced quantum applications.
\begin{figure}[h]
    \centering
    \includegraphics[width=1.0\linewidth]{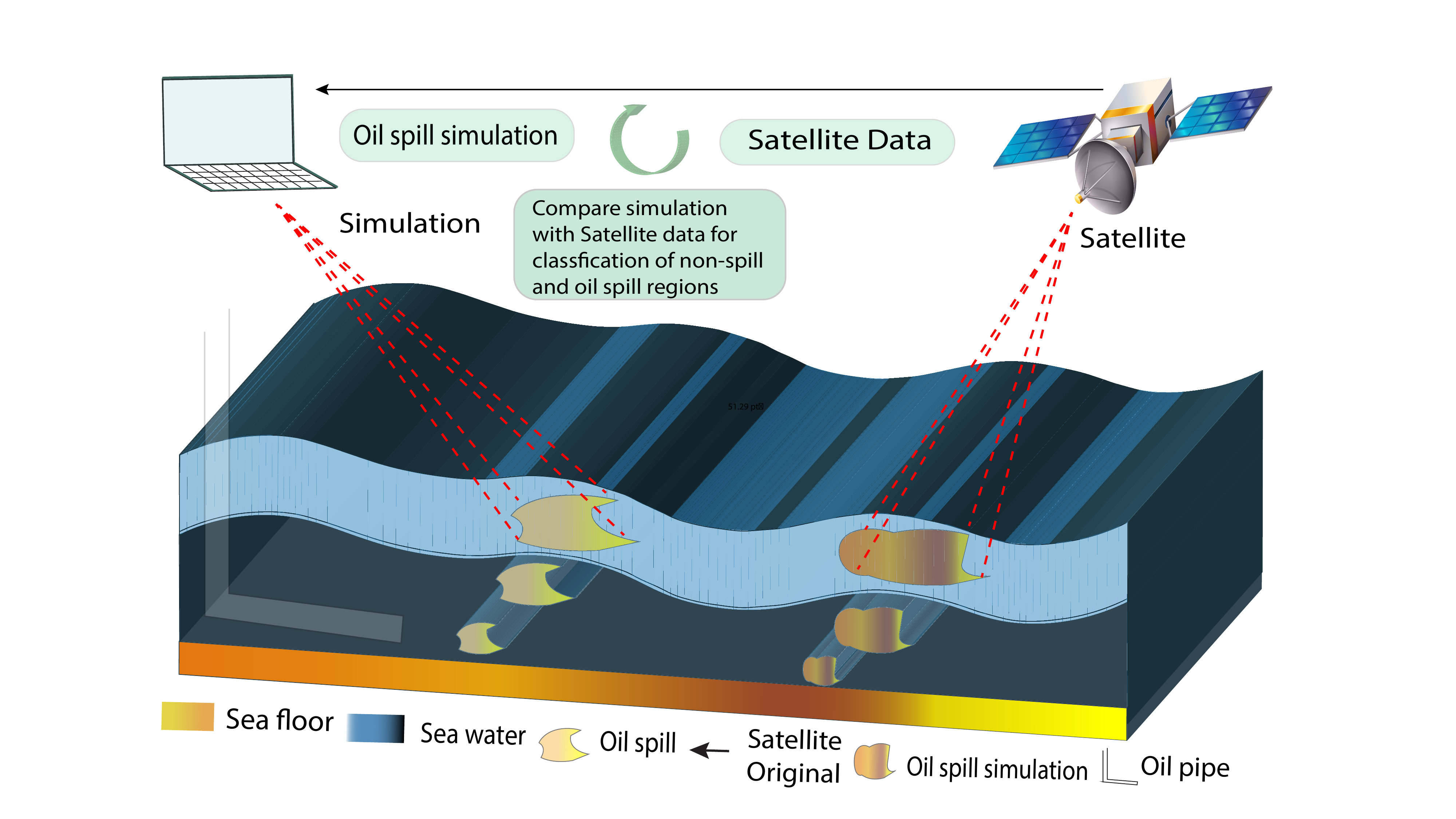}
    \vspace{-0.5cm}
    \caption{
   Overview of the simulation process that compares satellite imagery data to identify and classify non-spill and oil-spill regions. The simulation, performed on both QC devices and non-quantum devices, mimics oil-spill patterns, and these simulated images are cross-referenced with satellite data and extracted as valuable information for enhanced classification accuracy. The process focuses on refining the detection and differentiation between oil-spill-affected areas and clean regions for precise environmental analysis.}
    \label{satellite}
\end{figure}

This paper introduces a novel method using QBNs to classify imbalanced datasets into ``oil-spill'' and ``non-spill'' categories. The QBNs model employs quantum state preparation and probabilistic reasoning to predict oil-spills from satellite data accurately, as shown in Fig. \ref{satellite}. Integrating the QBNs with classical ML models signifies a considerable advancement in environmental science, offering innovative solutions for complex detection challenges.

\begin{figure}[htpb]
    \centering
    \vspace{-1cm}
    \includegraphics[width=1.\linewidth]{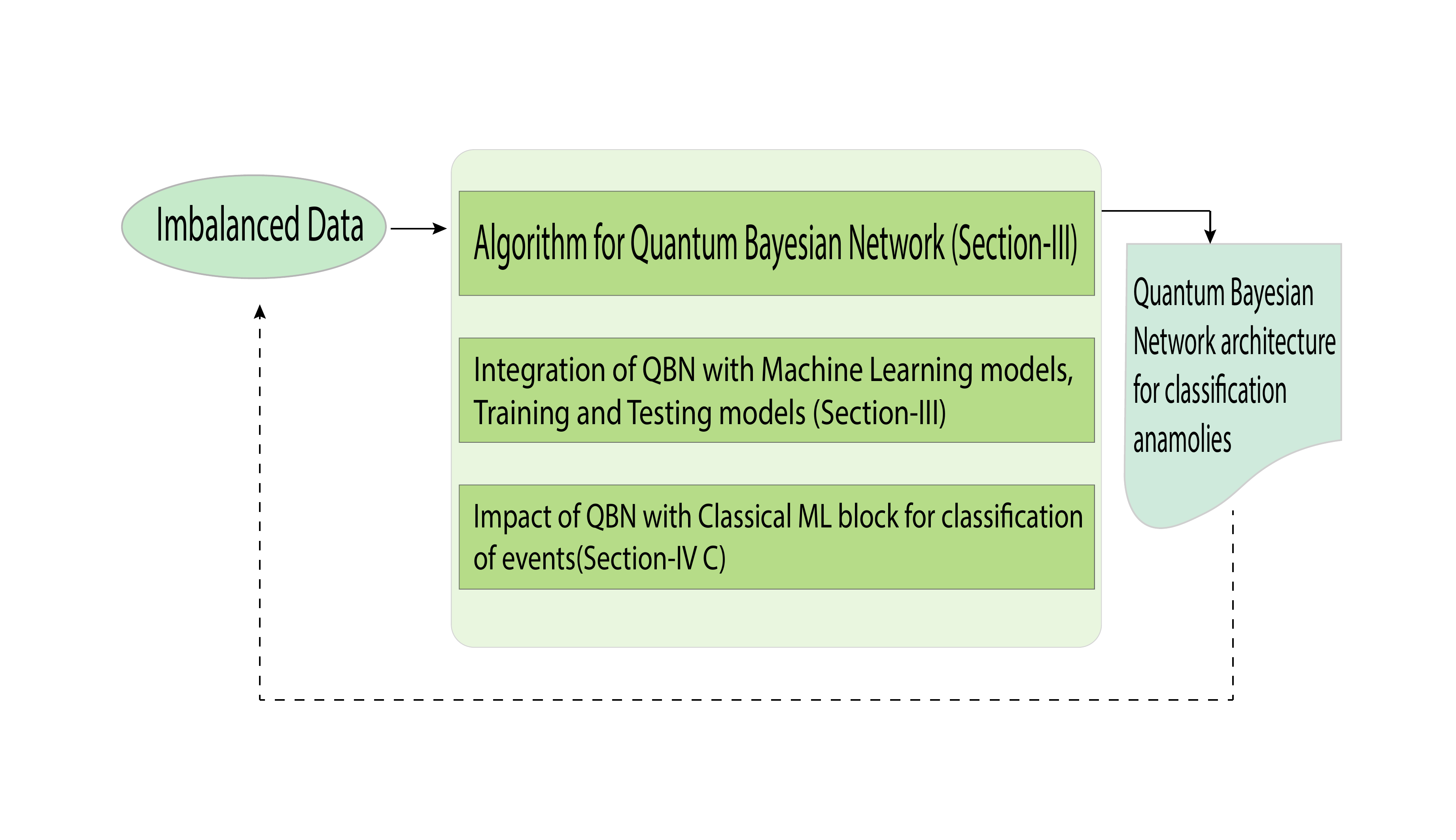}
    \vspace{-1cm}
    \caption{A conceptual representation of our contributions for imbalanced classification, illustrating the development of QBNs algorithm, its integration with machine learning models, and their impact on anomaly classification.}
    \label{novel}
\end{figure}

\textbf{Our contributions are outlined as follows} (see  Fig. \ref{novel}):
\begin{itemize}
    \item We develop a QBNs algorithm specifically optimized to enhance the detection of oil-spills using satellite imagery. 

    \item We design the QBNs algorithm to employ controlled parameterized gates within quantum circuits for calculating conditional probabilities based on extracted features. We further introduce a sensitivity threshold to enhance the detection of subtle variations, improving feature resolution and classification accuracy in imbalanced datasets.

    \item We integrate QBNs with six classical ML models to demonstrate their combined effectiveness in detecting oil spills, achieving performance improvements in several hybrid setups.

    \item We map the QBNs algorithm to IBM's 127-qubit quantum processor, IBM-Kyiv, and conduct transpilation across different optimization levels to evaluate its performance under varying conditions.

    \item
We demonstrate strong classification performance across classical and hybrid settings, achieving an Area Under the Curve (AUC) score of 1.00 with classical ML models and 0.99 when integrating QBNs with ML models.

\end{itemize}
\section{Classical Bayesian Networks}
Bayesian Networks are probabilistic models used to solve problems with associated uncertainties \cite{carrascal2023bayesian}. They are capable of representing uncertainties in the data and incorporating expert knowledge into the model \cite{Vicente}. Each feature or node represents a random variable, and the strength of the relationship between variables is quantified by the conditional probabilities associated with each feature.

The Bayesian network functions to compute the posterior distribution of a set of features, given that other feature values constitute observations or evidence. This step updates the probabilities and represents the flow of information throughout the network. To update the information, it uses the application of Bayes' theorem \cite{N.Innan}.
\begin{figure}[htpb]
    \centering
    \vspace{-1cm}
    \includegraphics[width=1\linewidth]{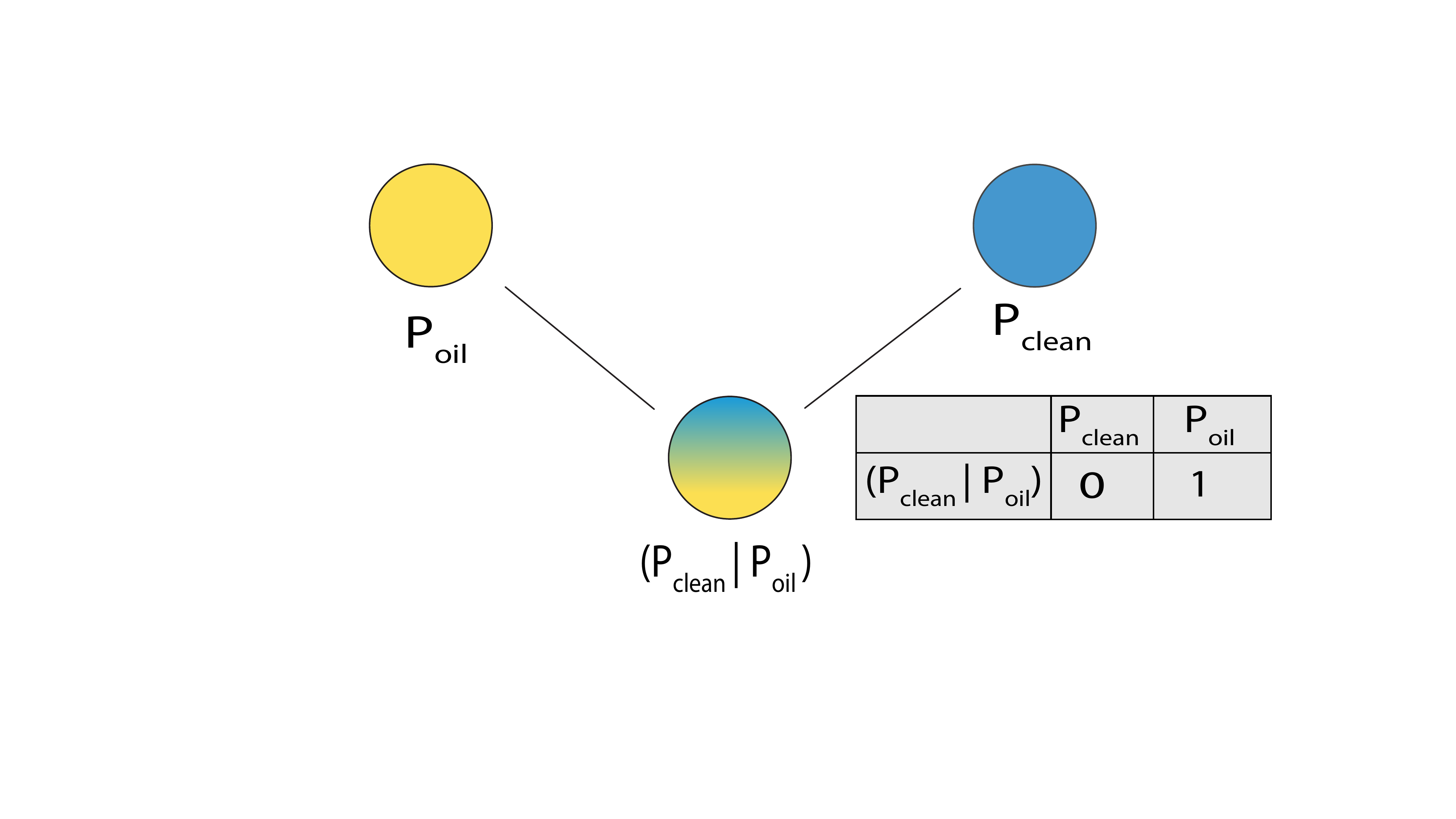}
    \vspace{-1.4cm}
    \caption{Structure of the Bayesian Networks for oil-spill and non-spill Regions. The diagram illustrates the Bayesian networks, where yellow nodes represent the oil-spill region $P_{\text{oil}}$ and blue nodes represent the non-spill region $P_{\text{non-spill}}$. The central node describes the conditional probability $\left(P_{\text{non-spill}}|P_{\text{oil-spill}}\right)$, indicating how the presence of oil influences the classification of non-spill patches. The accompanying table shows the conditional probabilities, where the values reflect the relationship between non-spill and oil-spill regions. 
    }
    \label{BN}
\end{figure}

For example, the two events are occurring as shown in Fig. \ref{BN}, oil-spills and non-spill water, such that the probability of both events is not zero ($P(\text{oil-spill}) \neq 0$, $P(\text{non-spill}) \neq 0$). We can express the conditional probability of an oil-spill given non-spill water as follows:
\begin{equation}
    P(\text{oil-spill} \mid \text{non-spill}) = \frac{P(\text{non-spill} \mid \text{oil-spill}) \cdot P(\text{oil-spill})}{P(\text{non-spill})},
\end{equation}
Additionally, for $n$ events such that $P(\text{oil-spill}) \neq 0$ for all $i$ within the range $1 \leq i \leq n$, the conditional probability can be generalized. Specifically, the probability of the $i$-th event $P(\text{oil-spill}_i \mid \text{non-spill})$ occurring given non-spill water can be written as:
\begin{equation}
\frac{P(\text{non-spill} \mid \text{oil-spill}_i) \cdot P(\text{oil-spill}_i)} 
     {\sum_{j=1}^{n} P(\text{non-spill} \mid \text{oil-spill}_j) \cdot P(\text{oil-spill}_j)}.
\end{equation}

For large datasets, the computational complexity also increases, this escalation directly impacts the time and memory requirements \cite{Marco}. These scaling challenges make efficient computation infeasible beyond a certain threshold. Quantum technologies, however, offer the potential to mitigate these limitations by employing quantum principles, providing a promising alternative approach to addressing certain complex problems. 

\begin{figure*}[t]
    \centering
    \includegraphics[width=1.0\linewidth]{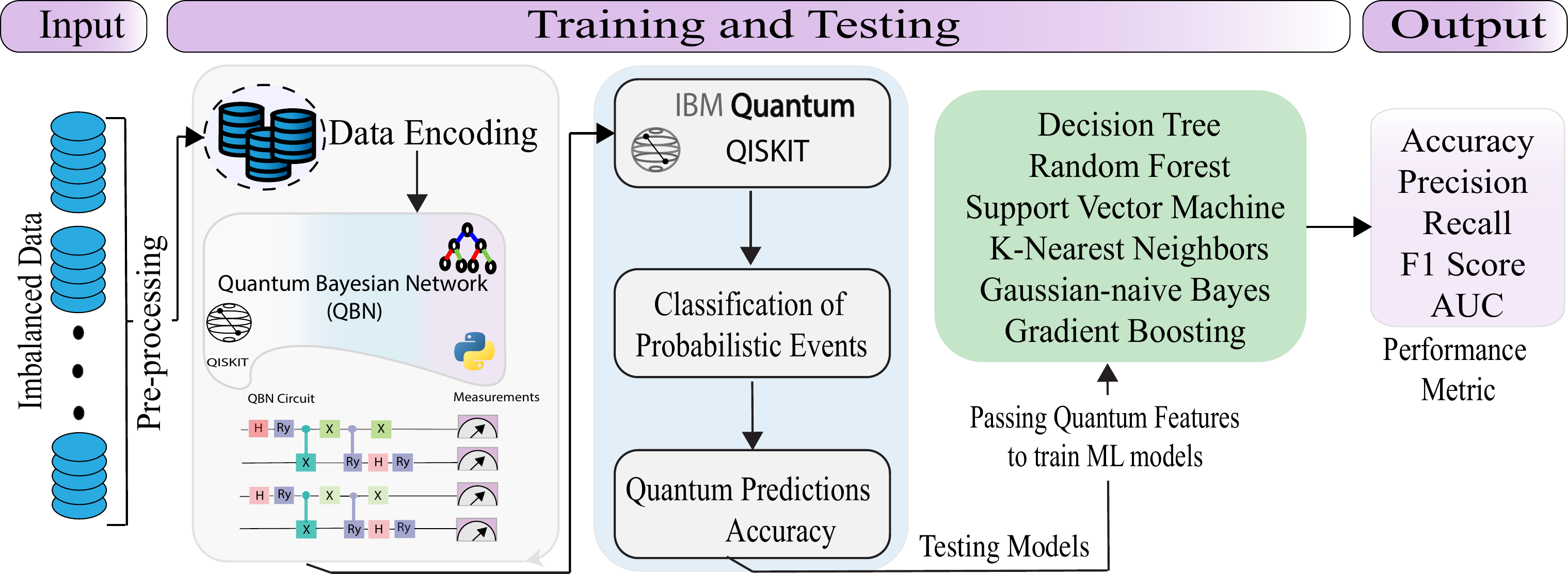}
    
    \caption{ Overall framework for QBNs classification. The framework starts with the input of imbalanced data, which is then processed by the QBNs using Qiskit for quantum state preparation and classification of probabilistic events. The quantum predictions are assessed for accuracy, and the resulting quantum features are used to train and test various classical ML models. The output performance metrics include accuracy, precision, recall, AUC, and F1 score.}
    \label{QBOCHS}
\end{figure*}

\section{Quantum Bayesian Networks}
In this section, we introduce the QBNs framework (see Fig. \ref{QBOCHS}), which integrates the quantum computational capabilities of QML. The proposed architecture is designed to accurately predict anomalies in imbalanced datasets. Like other QML models, it begins with data encoding, followed by a sequence of quantum operations that form the core of the processing circuit. The model predicts classes by performing measurements and sampling the probabilities. Based on these predictions, the accuracy is computed.

QBNs operate on unstructured data for $\mathbf{N}$-qubits. In this work, QBNs are employed to analyze the differences between the two classes. The data is provided such that each dataset is mapped to a qubit, and quantum gate operations are performed to determine the correct class. The QBNs undergo the following steps:
\begin{itemize} 
\item \textbf{Data Preparation:} Preparation of the dataset and splitting it into features and true labels. 
\item \textbf{Quantum Circuit:} 
\begin{enumerate} 
\item Initialize a quantum circuit with \textit{N}-qubits corresponding to the number of features ($N$) as presented in Fig. \ref{QBNcircuit}. 

\item Incorporate quantum gate operations to compute the probabilities of the two classes, representing a probabilistic model where each qubit's state corresponds to a random variable in the Bayesian network. \end{enumerate} 
\begin{figure}[h]
    \centering
    \vspace{-0.8cm}   
    \includegraphics[width=1\linewidth]{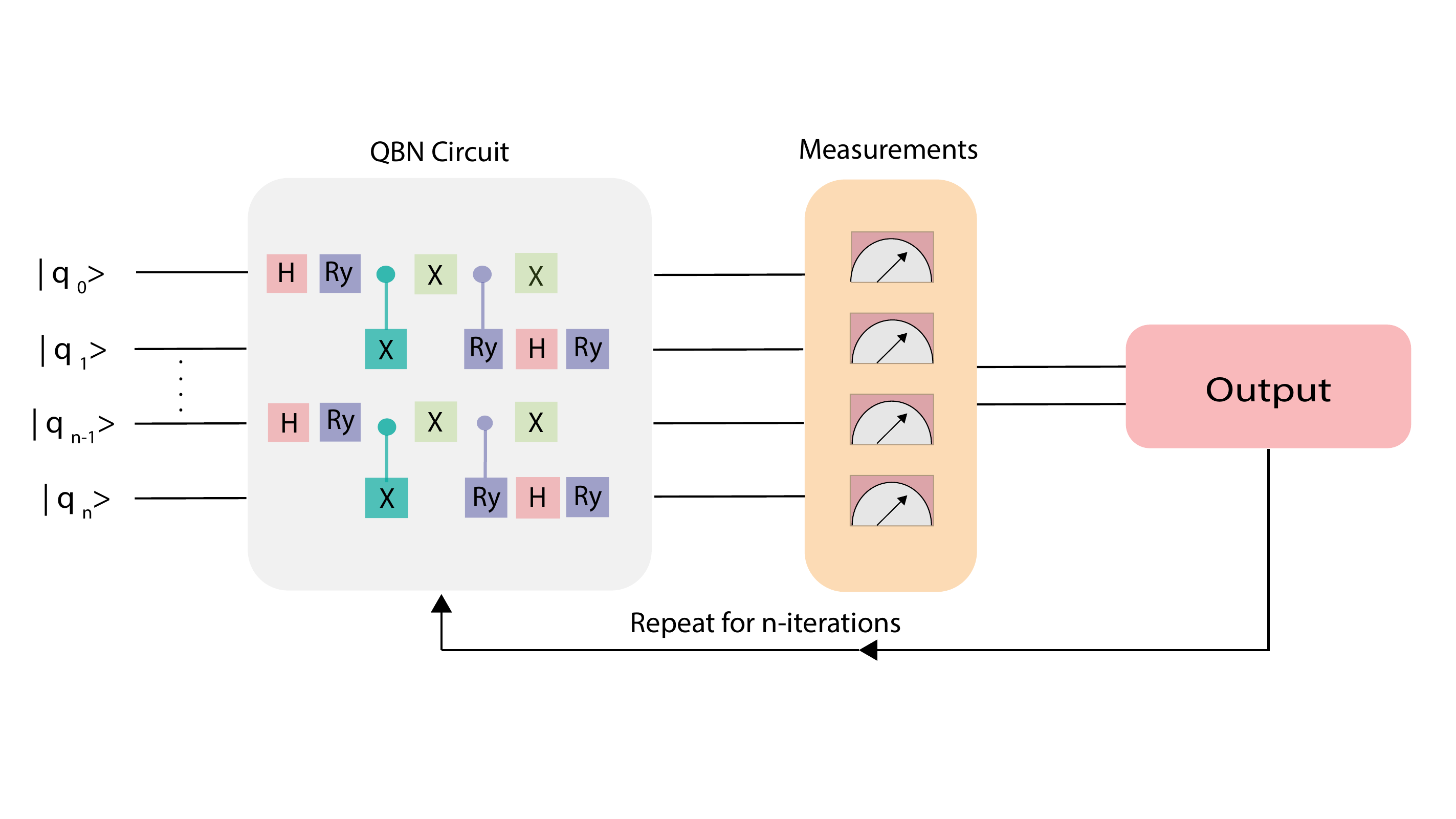}
    \vspace{
    -1.4cm}
    \caption{The QBNs circuit. It starts with Hadamard (H) gates to create superposition states, followed by Rotation-Y (RY) gates to encode data, and CNOT gates for entanglement. Measurements are taken, and the process is repeated for $n$ iterations to produce the output probabilities for classification.}
    \label{QBNcircuit}
\end{figure}

\item \textbf{Quantum Operations:} 
To perform quantum operations on qubits, we apply superposition and parameterized quantum gates based on feature value, as described in Algorithm \ref{alg:QBN}.\\
\item \textbf{Training QBNs:} 
\begin{enumerate} 
\item Execute the quantum circuit on a quantum simulator. 
\item Obtain results in terms of counts. 
\end{enumerate}
\end{itemize}

Overall, the quantum circuit for QBNs operates by evolving the quantum state according to the encoded probabilities. When quantum states are measured, the circuit outputs a probability distribution over all possible outcomes, which can then be used for classification. The accuracy of the model is determined by comparing the quantum-derived predictions with the true labels to verify the model's performance.
\begin{algorithm}[h]
\DontPrintSemicolon
\caption{QBNs}
\label{alg:QBN}
\textbf{Input:} $N-$qubits, feature values $f_i$ for $i=1,\ldots,n$, maximum feature value $f_{\text{max}}$\;
\textbf{Output:} Quantum state encoding the feature values and their conditional probabilities\;

\textbf{Initialize each qubit to $|0\rangle$}\;
\For{$i = 1$ \textbf{to} $n$}{
    Apply Hadamard gate $H$ to qubit $i$: $H_i$\;
}
\For{$i = 1$ \textbf{to} $n$}{
    Calculate rotation angle $\theta_i = \frac{\pi}{4} \left( \frac{f_i}{f_{\text{max}}} \right)$\;
    Apply parameterized rotation gate $RY(\theta_i)$ to qubit $i$\;
}
\For{$i = 1$ \textbf{to} $n-1$}{
    Apply Controlled-RY gate $CU^{rot}_{i,i+1}(\theta_i)$ for entanglement between qubits $i$ and $i+1$\;
}
\For{$i = 1$ \textbf{to} $n-1$}{
    Apply CNOT gate between qubits $i$ and $i+1$ to further entangle them: $CNOT_{i,i+1}$\;
}
\textbf{Measure the quantum state}\;
\end{algorithm}

\section{Results and Discussion}

\subsection{Experimental Setup}

\begin{figure}[htpb]
    \centering
    \includegraphics[width=1.0\linewidth]{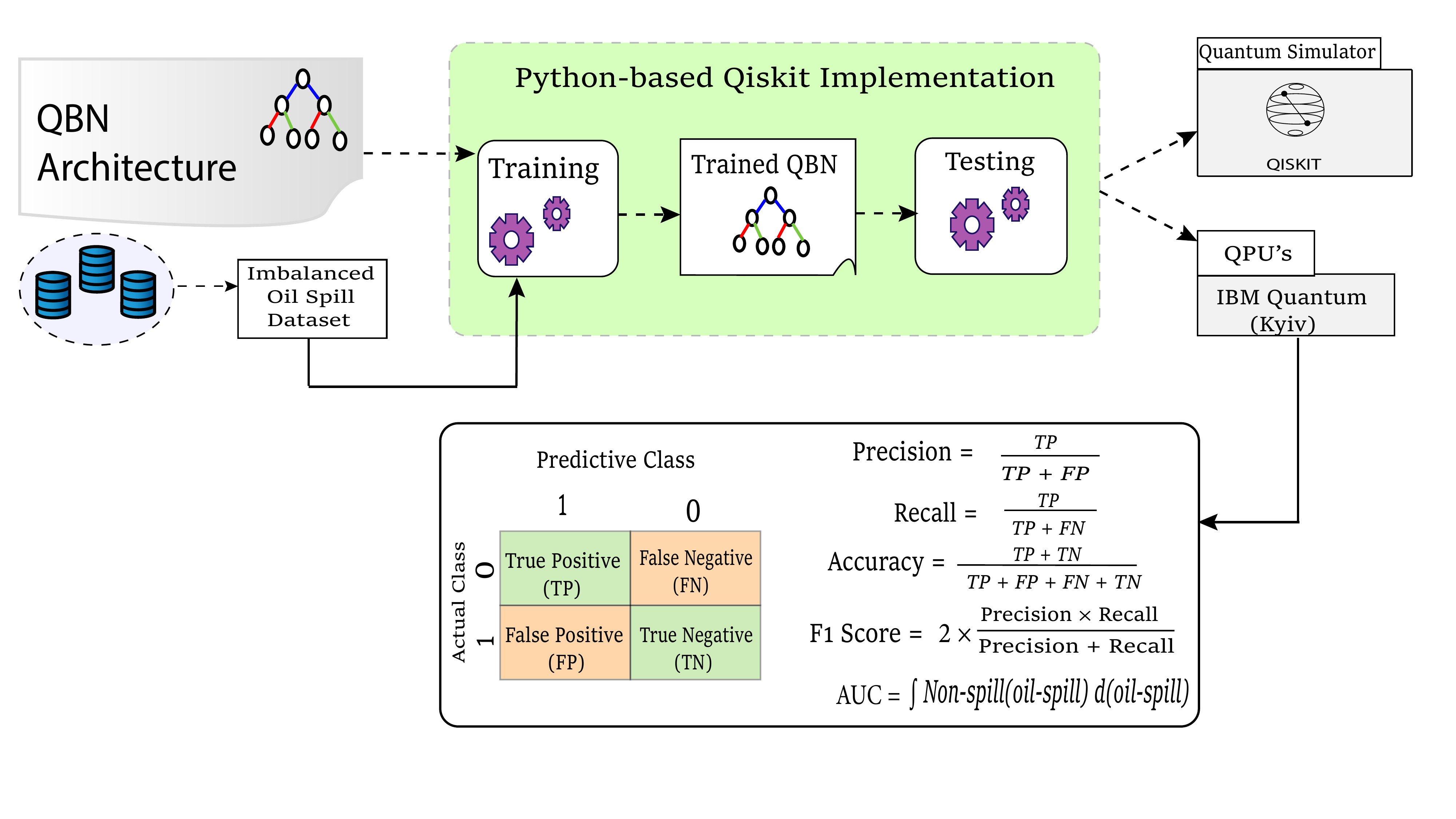}
    \vspace{-0.8cm}
    \caption{Experimental setup of the proposed approach, the imbalanced dataset is processed within the QBNs framework, implemented using Qiskit. The framework undergoes training on both a quantum simulator and IBM QPUs. Model performance is evaluated using standard binary classification metrics, including precision, recall, accuracy, and F1 score.}
    \label{Fig 9}
\end{figure}

The dataset used in this study originates from satellite images of the ocean \cite{data}, including both images with and without oil-spills. The images are segmented into smaller sections and processed using computer vision algorithms to yield a vector of features describing each image section, or patch. The primary task is to predict whether a given vector indicates the presence of an oil-spill, which is essential for identifying spills from illegal or accidental dumping in the ocean. The dataset consists of two classes: non-spill (clean), the majority class representing ocean patches without oil-spills, and oil-spill, the minority class representing patches with oil-spills. The objective is to distinguish between these classes using the extracted features. 
For the experimental framework, 20\% of the dataset is reserved for testing purposes, with the number of qubits $=48$ in the quantum circuit, set equal to the number of features to maximize computational efficiency. Fig. \ref{Fig 9} shows the structured framework for assessing our proposed methodology. The setup is run in the Python-based Qiskit environment. We employ the Qiskit Aer-Simulator to train our QBNs, as well as one of IBM's quantum hardware devices \cite{javadi2024quantum}, by performing multiple iterations to discern between the non-spill and oil-spill classes effectively.

\subsection{Training and Testing}
The quantum circuit for the Bayesian network is trained on the dataset, mapping features onto $N$-qubits, followed by performing quantum operations. 
The quantum simulator runs the circuit, whereas the quantum circuit predicts outcomes based on the measurements. We assign $0$ for non-spill samples and $1$ for oil-spill samples. 
For each feature, the normalized quantum states corresponding to non-spill and oil-spill events are defined as:
\begin{equation}
    |\Psi_{\text{non-spill}}\rangle = \frac{|\psi_{\text{non-spill}}\rangle}{\sqrt{|\langle \psi_{\text{non-spill}} | \psi_{\text{non-spill}} \rangle|^{2} + |\langle \psi_{\text{oil-spill}} | \psi_{\text{oil-spill}} \rangle|^{2}}},
\end{equation}   
\begin{equation}
    |\Psi_{\text{oil-spill}}\rangle = \frac{|\psi_{\text{oil-spill}}\rangle}{\sqrt{|\langle \psi_{\text{non-spill}} | \psi_{\text{non-spill}} \rangle|^{2} + |\langle \psi_{\text{oil-spill}} | \psi_{\text{oil-spill}} \rangle|^{2}}}, 
\end{equation}\\
where $|\psi_{\text{non-spill}}\rangle$ and $|\psi_{\text{oil-spill}}\rangle$, represent the non-normalized quantum states for non-spill and oil-spill events, respectively.

To further assess performance, ML models undergo training and testing, including Decision Tree (DT), Support Vector Machine (SVM), Random Forest (RF), K-Nearest Neighbors (KNN), Gaussian Naive Bayes (GnB), and Gradient Boosting (GB).
\subsection{Performance Metrics}
\begin{itemize}
    \item \textbf{Accuracy:} This metric evaluates the proportion of correct predictions relative to the total predictions made by the model, it can be defined as: 
\begin{equation}
    \text{Accuracy} = \frac{\sum_{i=1}^{N}\left(\hat{y_{i}}=y_i\right)}{N}.
\end{equation}
This metric calculates the effectiveness of QBN in classifying oil-spill and non-spill regions.
\item \textbf{Precision:} We evaluate precision as correctly predicted non-spill (True Positives ``TP'') and oil-spill (False Positives ``FP'') samples,
\begin{equation}
    \text{Precision} = \frac{\text{TP}}{\text{TP + FP}}.
\end{equation}
\item \textbf{Recall:} Correctly predicted positive classes to the total of correctly predicted TP and False Negative (FN),
\begin{equation}
    \text{Recall} = \frac{\text{TP}}{\text{TP + FN}}.
\end{equation}
\item \textbf{F1 Score:} Harmonic mean of precision and recall,
\begin{equation}
    \text{F1} = 2 \times \frac{\text{Precision} \times \text{Recall}}{\text{Precision + Recall}}
\end{equation}
\item \textbf{AUC:} It is computed based on the True Positive Rate (TPR) plotted against the False Positive Rate (FPR) at various threshold settings.
\begin{equation}
    AUC = \int_{0}^{1} Non-spill (oil-spill) d(oil-spill).
\end{equation}
\end{itemize}
\begin{figure*}[htbp]
    \centering
\includegraphics[width=1.0\linewidth]{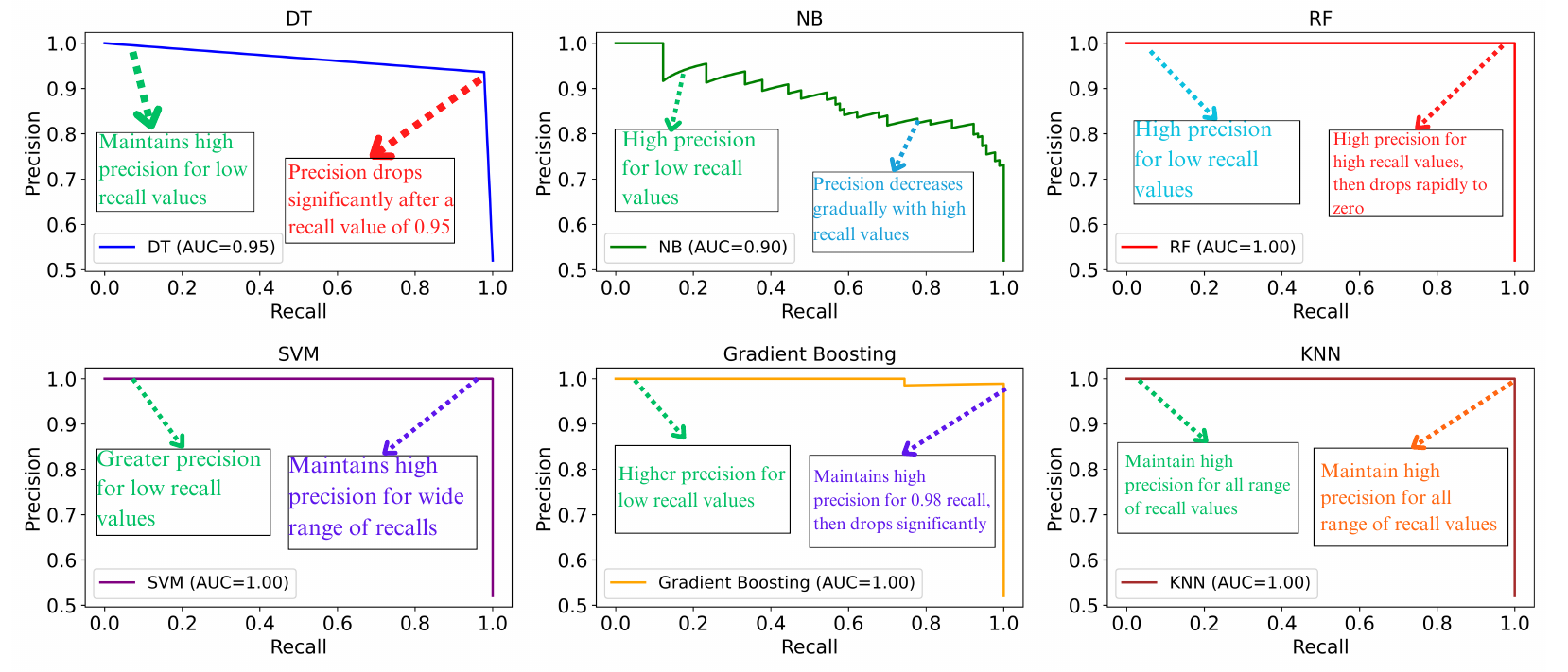}
    \caption{Precision-recall curves for classical ML models. The plots are arranged from top-left to bottom-right, indicating the performance metrics with recall on the x-axis and precision on the y-axis, exploiting the effectiveness of each model.}
    \vspace{-0.3cm}
    \label{classical-AUC}
\end{figure*}
\subsection{Experimental Results}
Our research begins with evaluating classical ML models, as presented in Table \ref{report1}. During testing, precision-recall curves, which are plotted in Fig. \ref{classical-AUC}, help assess the capability of these models with a specific emphasis on the AUC score. The performance of these models varies, with RF, SVM, KNN, and GB models showing the most favorable outcomes, while GnB has the lowest performance in an imbalanced dataset.
These precision-recall curves represent the trade-offs involved in predicting minor classes, with SVM and KNN exhibiting the highest overall performance among the ML models. In contrast, GB remains the second-best performer compared to DT, which shows good precision at higher recall levels (AUC=0.95). RF displays an excellent overall performance (AUC=1.00), and GnB shows a gradual decrease in performance as recall levels increase (AUC=0.90). GB maintains high precision throughout (AUC=1.00), outperforming DT.
\begin{table}[htpb]
    \caption{Performance metrics for ML models show consistently high precision, recall, F1-score, and accuracy, indicating excellent classification performance across all models. AUC values vary, with RF, SVM, KNN, and GB achieving the highest, while DT and GnB are the lowest.}
    \label{report1}
    \centering
\begin{adjustbox}{max width=\linewidth}
\begin{tabular}{@{}lcccccc@{}}
\toprule
  \textbf{Models} & \textbf{Precision} & \textbf{Recall} &  \textbf{F1-Score} & \textbf{Accuracy} &  \textbf{AUC}\\ \midrule
DT & 0.95  & 0.93  & 0.95 & 0.95 & 0.95\\
  RF & 1.00  & 1.00   & 1.00 & 1.00 & 1.00\\
  SVM & 1.00  & 1.00   & 1.00 & 1.00 & 1.00\\
  KNN & 1.00  & 1.00   & 1.00 & 1.00 & 1.00\\
  GnB & 1.00  & 0.42   & 0.59 & 0.72 & 0.90\\
  GB & 1.00  & 0.94   & 0.97 & 0.97 & 1.00\\
\bottomrule
\end{tabular}
\end{adjustbox}
\end{table}

\begin{figure}[htpbt]
    \centering
    \includegraphics[width=1.0\linewidth]{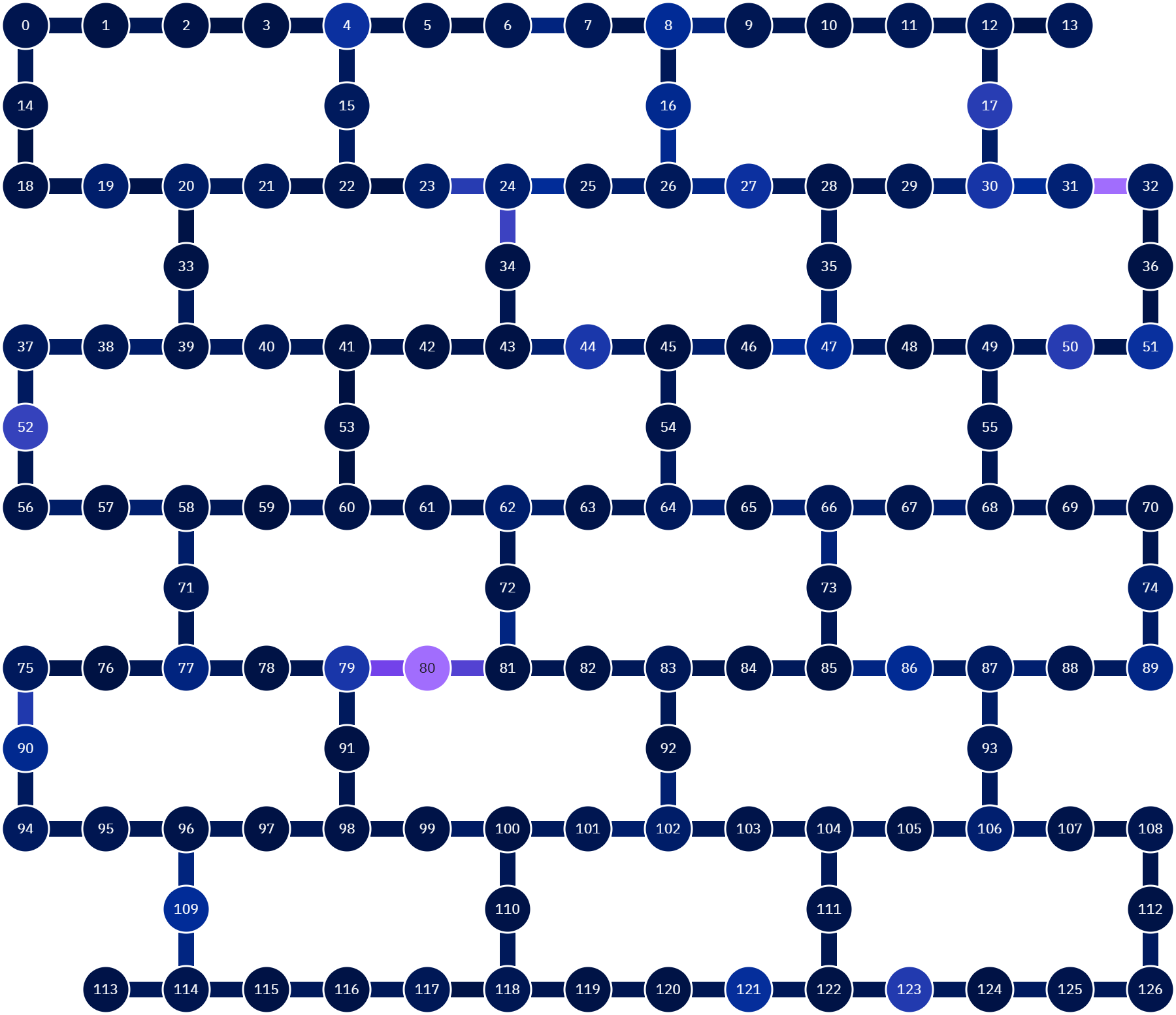}
    \vspace{-0.15cm}
    \caption{IBM-Kyiv quantum hardware architecture calibration read-outs.}
    \label{IBM-Kyiv}
\end{figure}
Before integrating our QBNs framework with classical ML models, we evaluate its standalone classification performance on the imbalanced dataset. QBNs function as a classifier, achieving consistently high precision, recall, and F1-scores near 0.97 across various test scenarios, as shown in Table \ref{qbns1}. These results demonstrate the effectiveness of QBNs in handling complex data distributions and imbalances. For the quantum hardware evaluation, we deploy QBNs on IBM-Kyiv quantum hardware, which features a 127-qubit architecture, as shown in Fig. \ref{IBM-Kyiv}. 

The quantum circuits are transpiled with optimization levels 2 and 3, where level 2 incorporates noise-adaptive qubit mapping and level 3 includes gate cancellation and unitary synthesis. Optimization improves QBNs' metrics, increasing precision, recall, and F1-scores by approximately 0.01\%, while the AUC remains unchanged due to the dataset's inherent imbalance. The low AUC results from the disproportionate ratio of TP to FN, which affects the model's ability to balance performance across all classes.
\begin{figure*}[b]
    \centering
\includegraphics[width=1.0\linewidth]{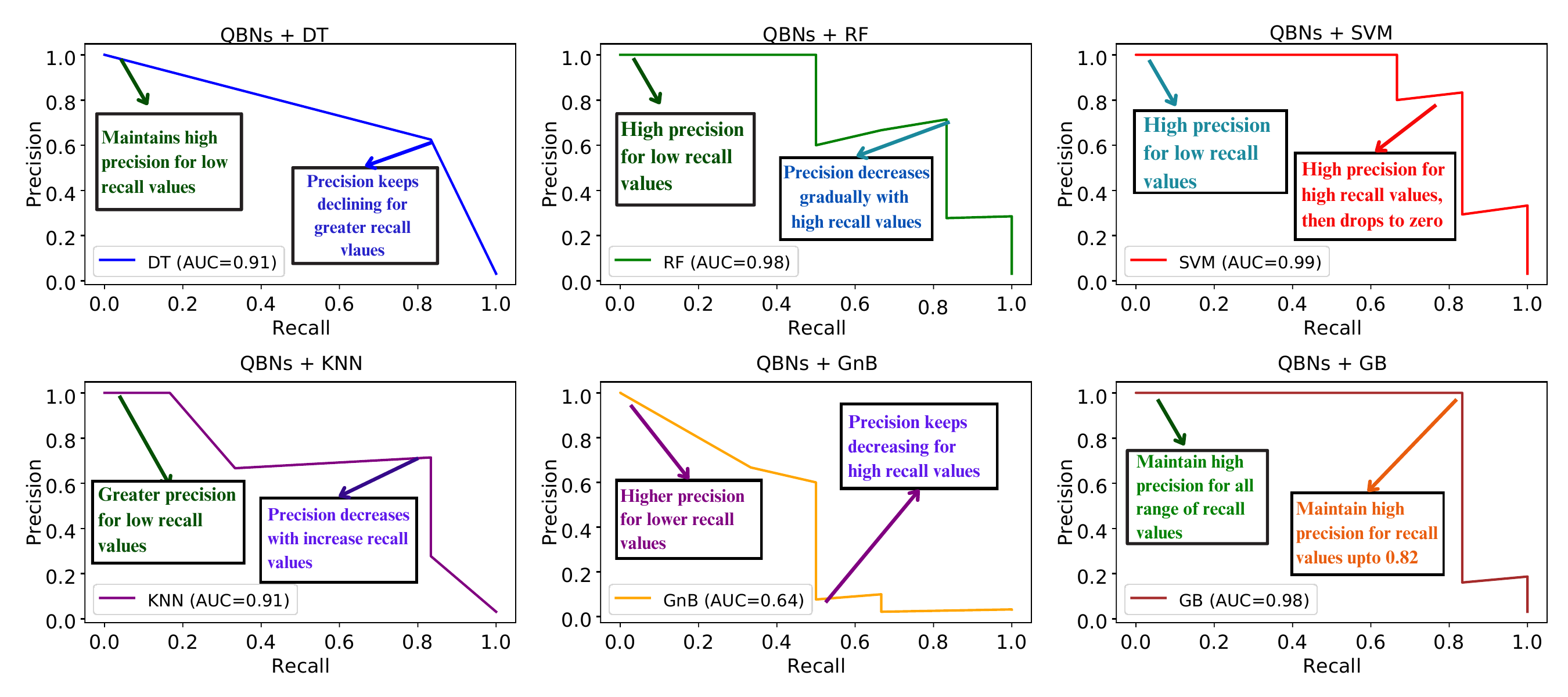}
    \caption{Precision-recall curves for various ML models integrated with QBNs. The plots, arranged from top-left to bottom-right, depict the performance metrics with recall on the x-axis and precision on the y-axis, demonstrating the effectiveness of each model combination.}
    \vspace{-0.3cm}
    \label{subplots}
\end{figure*}
To address these challenges and further enhance classification performance, we integrate QBNs with standard ML models. As shown in Table \ref{report}, this hybrid approach consistently achieves precision, recall, and F1-scores around 0.99 across most models, with notable improvements in AUC for combinations with RF, SVM, and GB. Precision-recall curves in Fig. \ref{subplots} provide additional insights into the trade-offs between precision and recall at varying thresholds. For instance, QBNs combined with RF and SVM maintain high precision across most levels of recall, achieving AUC values of 0.97 and 0.99, respectively. However, QBNs with GnB show lower AUC values (0.64), likely due to misaligned probabilistic dependencies in the imbalanced dataset. This highlights the importance of considering class dependencies and feature relationships when integrating QBNs with specific ML models.

\begin{table}[!ht]
\caption{Performance metrics for QBNs.}
\centering
\begin{adjustbox}{max width=\linewidth}
\begin{tabular}{@{}llcccccc@{}}
\toprule
\multicolumn{2}{c}{\textbf{Backend}} & \textbf{Precision} & \textbf{Recall} & \textbf{F1-Score} & \textbf{Accuracy} & \textbf{AUC}
   \\ \midrule
\multicolumn{2}{c}{AerSimulator} & 0.96               & 0.99             & 0.97               & 0.95   & 0.58  \\          
\multirow{2}{*}{IBM-Kyiv} & (Opt. level = 2)   & 0.97 \textuparrow    & 0.99  \textuparrow            & 0.98 \textuparrow    & 0.96 \textuparrow  & 0.58 \\
& (Opt. level = 3)   & 0.97 \textuparrow    & 1.00  \textuparrow            & 0.99 \textuparrow    & 0.97 \textuparrow  & 0.58 \\
\bottomrule
\end{tabular}
\end{adjustbox}
\label{qbns1}
\end{table}

\begin{table}[htpb]
\centering

\caption{Performance metrics for ML models enhanced by QBNs show consistently high precision, recall, F1-score, and accuracy, indicating excellent classification performance across all models. AUC values vary, with QBNs+RF, QBNs+SVM, and QBNs+GB achieving the highest and QBNs+GnB the lowest.}
\label{report}
\begin{adjustbox}{max width=\linewidth}
\begin{tabular}{@{}lcccccc@{}}
\toprule
  \textbf{Models} & \textbf{Precision} & \textbf{Recall} &  \textbf{F1-Score} & \textbf{Accuracy} &  \textbf{AUC}\\ \midrule
  QBNs+DT &  0.99 \textdownarrow & 0.99 \textuparrow   & 0.99 \textuparrow & 0.99 \textuparrow & 0.91 \textdownarrow\\
  QBNs+RF & 0.99  \textdownarrow & 0.99  \textdownarrow & 0.99 \textdownarrow & 0.99\textdownarrow & 0.98 \textdownarrow\\
  QBNs+SVM & 0.99 \textdownarrow & 0.99 \textdownarrow  & 0.99 \textdownarrow & 0.99 \textdownarrow & 0.99 \textdownarrow\\
  QBNs+KNN & 0.99 \textdownarrow & 0.99 \textdownarrow   & 0.99 \textdownarrow & 0.99 \textdownarrow & 0.91 \textdownarrow\\
  QBNs+GnB & 0.99 \textdownarrow & 0.99 \textuparrow   & 0.99 \textuparrow & 0.99 \textuparrow & 0.64 \textdownarrow\\
  QBNs+GB & 0.99 \textdownarrow & 0.99 \textuparrow   & 0.99 \textuparrow & 0.99 \textuparrow & 0.98 \textdownarrow\\
\bottomrule
\end{tabular}
\end{adjustbox}
\end{table}

\subsection{Discussion}
Integrating QBNs with ML models marks a significant step forward in applying QC to environmental monitoring. This research highlights the effectiveness of QBNs in managing complex and imbalanced dataset, commonly found in situations like oil spill detection over vast ocean areas.  By rigorously validating QBNs algorithm performance, we demonstrate its ability to enhance the discriminative power and accuracy of standard ML models, as reflected in improved AUC scores.

The hybrid integration of QC techniques with classical ML models capitalizes on the strengths of both approaches while mitigating their individual limitations. Our standalone evaluation of QBNs on the IBM-Kyiv quantum hardware provides valuable insights into their capabilities and limitations as a classifier, highlighting the necessity of integrating them within a hybrid strategy to further enhance performance.

This hybrid computational approach employs the processing capabilities of quantum mechanisms and the robustness of classical ML algorithms, resulting in a more effective framework for monitoring and managing environmental hazards. Moreover, the practical application of QBNs in this context demonstrates the feasibility and utility of quantum-enhanced ML models in real-world scenarios beyond theoretical or experimental boundaries. For tasks like distinguishing between oil-spill and non-spill events from satellite imagery, QBNs prove to be a viable and beneficial solution.

Our findings further reveal that certain classical ML models benefit significantly from quantum integration. This nuanced understanding of each model's strengths and limitations suggests that tailored approaches may be necessary to optimize hybrid quantum-classical systems for specific environmental applications. 

\section{Conclusion}
Our study introduces a novel quantum Bayesian approach for oil-spill detection, employing QBNs to accurately classify two distinct categories: non-spill and oil-spill. We achieve this by initializing a quantum circuit with $N-$qubits corresponding to each feature, processing data from an imbalanced dataset over n-iterations.
This framework facilitates precise feature extraction and enables exact data categorization. Parameters are carefully optimized to train classical ML models on the outputs from the QBNs, allowing for performance validation through rigorous testing. The results are both promising and affirming, showcasing the QBNs framework's ability to detect oil-spills accurately, thus proving its practical applicability in real-world environmental monitoring scenarios.

Nevertheless, the study acknowledges certain limitations. While most models exhibit strong performance, some integrations yield lower AUC scores, indicating variability in the effectiveness of the quantum-classical hybrid setup. This underlines the need for careful model selection and optimization to ensure consistent robustness and reliability in diverse operational scenarios. These findings highlight the potential of the proposed framework while paving the way for future research to refine the integration of quantum and classical models, thereby broadening its applicability and enhancing its efficiency in addressing complex environmental challenges.

\section*{Acknowledgments}
This work was supported in part by the NYUAD Center for Quantum and Topological Systems (CQTS), funded by Tamkeen under the NYUAD Research Institute grant CG008.
\bibliographystyle{IEEEtran}

{\footnotesize\bibliography{refs}}

\begin{thebibliography}{10}
\providecommand{\url}[1]{#1}
\csname url@samestyle\endcsname
\providecommand{\newblock}{\relax}
\providecommand{\bibinfo}[2]{#2}
\providecommand{\BIBentrySTDinterwordspacing}{\spaceskip=0pt\relax}
\providecommand{\BIBentryALTinterwordstretchfactor}{4}
\providecommand{\BIBentryALTinterwordspacing}{\spaceskip=\fontdimen2\font plus
\BIBentryALTinterwordstretchfactor\fontdimen3\font minus \fontdimen4\font\relax}
\providecommand{\BIBforeignlanguage}[2]{{%
\expandafter\ifx\csname l@#1\endcsname\relax
\typeout{** WARNING: IEEEtran.bst: No hyphenation pattern has been}%
\typeout{** loaded for the language `#1'. Using the pattern for}%
\typeout{** the default language instead.}%
\else
\language=\csname l@#1\endcsname
\fi
#2}}
\providecommand{\BIBdecl}{\relax}
\BIBdecl

\bibitem{crecy2024mt}
S.~Crecy, R.~Yender, and J.~Rossetti, ``Mt princess empress oil spill response in the philippines: International coordination and communication,'' in \emph{International Oil Spill Conference Proceedings}, vol. 2024, no.~1, 2024.

\bibitem{nixon2016shoreline}
Z.~Nixon \emph{et~al.}, ``Shoreline oiling from the deepwater horizon oil spill,'' \emph{Marine pollution bulletin}, 2016.

\bibitem{hettithanthri2024review}
O.~Hettithanthri \emph{et~al.}, ``A review of oil spill dynamics: Statistics, impacts, countermeasures, and weathering behaviors,'' \emph{Asia-Pacific Journal of Chemical Engineering}, p. e3128, 2024.

\bibitem{biamonte2017quantum}
J.~Biamonte \emph{et~al.}, ``Quantum machine learning,'' \emph{Nature}, vol. 549, no. 7671, pp. 195--202, 2017.

\bibitem{zaman2023survey}
K.~Zaman, A.~Marchisio, M.~A. Hanif, and M.~Shafique, ``A survey on quantum machine learning: Current trends, challenges, opportunities, and the road ahead,'' \emph{arXiv preprint arXiv:2310.10315}, 2023.

\bibitem{rezaei2024environmental}
T.~Rezaei and A.~Javadi, ``Environmental impact assessment of ocean energy converters using quantum machine learning,'' \emph{Journal of Environmental Management}, 2024.

\bibitem{marchisio2024cutting}
A.~Marchisio, E.~Sychiuco, M.~Kashif, and M.~Shafique, ``Cutting is all you need: Execution of large-scale quantum neural networks on limited-qubit devices,'' \emph{arXiv preprint arXiv:2412.04844}, 2024.

\bibitem{kashif2022demonstrating}
M.~Kashif and S.~Al-Kuwari, ``Demonstrating quantum advantage in hybrid quantum neural networks for model capacity,'' in \emph{2022 IEEE international conference on rebooting computing (ICRC)}.\hskip 1em plus 0.5em minus 0.4em\relax IEEE, 2022, pp. 36--44.

\bibitem{kashif2024computational}
M.~Kashif, A.~Marchisio, and M.~Shafique, ``Computational advantage in hybrid quantum neural networks: Myth or reality?'' \emph{arXiv preprint arXiv:2412.04991}, 2024.

\bibitem{yamasaki2020learning}
H.~Yamasaki, S.~Subramanian, S.~Sonoda, and M.~Koashi, ``Learning with optimized random features: Exponential speedup by quantum machine learning without sparsity and low-rank assumptions,'' \emph{Advances in neural information processing systems}, 2020.

\bibitem{caro2022generalization}
M.~C. Caro \emph{et~al.}, ``Generalization in quantum machine learning from few training data,'' \emph{Nature communications}, vol.~13, no.~1, p. 4919, 2022.

\bibitem{Innan_Grover_2024}
N.~Innan and M.~Bennai, ``A variational quantum perceptron with grover’s algorithm for efficient classification,'' \emph{Physica Scripta}, vol.~99, no.~5, p. 055120, apr 2024.

\bibitem{innan2023enhancing}
N.~Innan, M.~A.-Z. Khan, B.~Panda, and M.~Bennai, ``Enhancing quantum support vector machines through variational kernel training,'' \emph{Quantum Information Processing}, vol.~22, no.~10, p. 374, 2023.

\bibitem{farhi2018classification}
E.~Farhi and H.~Neven, ``Classification with quantum neural networks on near term processors,'' \emph{arXiv preprint arXiv:1802.06002}, 2018.

\bibitem{maouaki2024designing}
W.~El~Maouaki, A.~Marchisio, T.~Said, M.~Shafique, and M.~Bennai, ``Designing robust quantum neural networks: Exploring expressibility, entanglement, and control rotation gate selection for enhanced quantum models,'' \emph{arXiv preprint arXiv:2411.11870}, 2024.

\bibitem{innan2024financial}
N.~Innan \emph{et~al.}, ``Financial fraud detection using quantum graph neural networks,'' \emph{Quantum Machine Intelligence}, vol.~6, no.~1, p.~7, 2024.

\bibitem{innan2024fedqnn}
N.~Innan, M.~A.-Z. Khan, A.~Marchisio, M.~Shafique, and M.~Bennai, ``Fedqnn: Federated learning using quantum neural networks,'' in \emph{International Joint Conference on Neural Networks (IJCNN)}, 2024.

\bibitem{el2024quantum}
W.~El~Maouaki, N.~Innan, A.~Marchisio, T.~Said, M.~Bennai, and M.~Shafique, ``Quantum clustering for cybersecurity,'' in \emph{International Conference on Quantum Computing and Engineering (QCE)}, 2024.

\bibitem{dutta2024qadqn}
S.~Dutta, N.~Innan, A.~Marchisio, S.~B. Yahia, and M.~Shafique, ``Qadqn: Quantum attention deep q-network for financial market prediction,'' in \emph{International Conference on Quantum Computing and Engineering (QCE)}, 2024.

\bibitem{innan2024qfnn}
N.~Innan, A.~Marchisio, M.~Shafique, and M.~Bennai, ``Qfnn-ffd: Quantum federated neural network for financial fraud detection,'' \emph{arXiv preprint arXiv:2404.02595}, 2024.

\bibitem{innan2024fl}
N.~Innan, A.~Marchisio, and M.~Shafique, ``Fl-qdsnns: Federated learning with quantum dynamic spiking neural networks,'' \emph{arXiv preprint arXiv:2412.02293}, 2024.

\bibitem{dutta2024mqfl}
S.~Dutta, N.~Innan, S.~B. Yahia, M.~Shafique, and D.~E.~B. Neira, ``Mqfl-fhe: Multimodal quantum federated learning framework with fully homomorphic encryption,'' \emph{arXiv preprint arXiv:2412.01858}, 2024.

\bibitem{maouaki2025qfal}
W.~E. Maouaki, N.~Innan, A.~Marchisio, T.~Said, M.~Bennai, and M.~Shafique, ``Qfal: Quantum federated adversarial learning,'' \emph{arXiv preprint arXiv:2502.21171}, 2025.

\bibitem{dutta2024federated}
S.~Dutta \emph{et~al.}, ``Federated learning with quantum computing and fully homomorphic encryption: A novel computing paradigm shift in privacy-preserving ml,'' \emph{arXiv preprint arXiv:2409.11430}, 2024.

\bibitem{chen2024crossing}
I.~Chen, N.~Innan, S.~K. Roy, and J.~Saroni, ``Crossing the gap using variational quantum eigensolver: A comparative study,'' \emph{arXiv preprint arXiv:2405.11687}, 2024.

\bibitem{innan2024quantum1}
N.~Innan, M.~A.-Z. Khan, and M.~Bennai, ``Quantum computing for electronic structure analysis: Ground state energy and molecular properties calculations,'' \emph{Materials Today Communications}, vol.~38, p. 107760, 2024.

\bibitem{innan2025qnn}
N.~Innan, B.~K. Behera, S.~Al-Kuwari, and A.~Farouk, ``Qnn-vrcs: A quantum neural network for vehicle road cooperation systems,'' \emph{IEEE Transactions on Intelligent Transportation Systems}, 2025.

\bibitem{innan2025optimizing}
N.~Innan, B.~K. Behera, S.~Mumtaz, S.~Al-Kuwari, A.~Farouk \emph{et~al.}, ``Optimizing low-energy carbon iiot systems with quantum algorithms: Performance evaluation and noise robustness,'' \emph{IEEE Internet of Things Journal}, 2025.

\bibitem{CerezoNature2022}
M.~Cerezo \emph{et~al.}, ``Challenges and opportunities in quantum machine learning,'' \emph{Nat Comput Sci}, 2022.

\bibitem{HuangHS2022}
H.-Y. Huang \emph{et~al.}, ``Quantum advantage in learning from experiments,'' \emph{Science}, vol. 376, no. 6598, pp. 1182--1186, 2022.

\bibitem{peral2024systematic}
D.~Peral-Garc{\'\i}a, J.~Cruz-Benito, and F.~J. Garc{\'\i}a-Pe{\~n}alvo, ``Systematic literature review: Quantum machine learning and its applications,'' \emph{Computer Science Review}, 2024.

\bibitem{liu2024laziness}
J.~Liu, Z.~Lin, and L.~Jiang, ``Laziness, barren plateau, and noises in machine learning,'' \emph{Machine Learning: Science and Technology}, vol.~5, no.~1, p. 015058, 2024.

\bibitem{el2024robqunns}
W.~El~Maouaki, A.~Marchisio, T.~Said, M.~Shafique, and M.~Bennai, ``Robqunns: A methodology for robust quanvolutional neural networks against adversarial attacks,'' in \emph{2024 IEEE International Conference on Image Processing Challenges and Workshops (ICIPCW)}.\hskip 1em plus 0.5em minus 0.4em\relax IEEE, 2024, pp. 4090--4095.

\bibitem{ahmed2025quantum}
T.~Ahmed, M.~Kashif, A.~Marchisio, and M.~Shafique, ``Quantum neural networks: A comparative analysis and noise robustness evaluation,'' \emph{arXiv preprint arXiv:2501.14412}, 2025.

\bibitem{el2024advqunn}
W.~El~Maouaki, A.~Marchisio, T.~Said, M.~Bennai, and M.~Shafique, ``Advqunn: A methodology for analyzing the adversarial robustness of quanvolutional neural networks,'' in \emph{International Conference on Quantum Software (QSW)}, 2024.

\bibitem{Alam2021}
M.~Alam, S.~Kundu, R.~O. Topaloglu, and S.~Ghosh, ``Quantum-classical hybrid machine learning for image classification (iccad special session paper),'' in \emph{ICCAD}, 2021.

\bibitem{tang2024review}
D.~Tang, Y.~Zhan, and F.~Yang, ``A review of machine learning for modeling air quality: Overlooked but important issues,'' \emph{Atmospheric Research}, p. 107261, 2024.

\bibitem{LITTLE2021105509}
D.~I. Little, S.~R. Sheppard, and D.~Hulme, ``A perspective on oil spills: What we should have learned about global warming,'' \emph{Ocean \& Coastal Management}, 2021.

\bibitem{SimaEQCE2020}
S.~E. Borujeni, N.~H. Nyugen, S.~Nannapaneni, E.~C. Behrman, and J.~E. Steck, ``Experimental evaluation of quantum bayesian networks on ibm qx hardware,'' \emph{QCE}, 2020.

\bibitem{wang2024quantum}
M.-M. Wang and X.-Y. Zhang, ``Quantum bayes classifiers and their application in image classification,'' \emph{Physical Review A}, vol. 110, no.~1, p. 012433, 2024.

\bibitem{borujeni2021quantum}
S.~E. Borujeni, S.~Nannapaneni, N.~H. Nguyen, E.~C. Behrman, and J.~E. Steck, ``Quantum circuit representation of bayesian networks,'' \emph{Expert Systems with Applications}, vol. 176, p. 114768, 2021.

\bibitem{harikrishnakumar2023forecasting}
R.~Harikrishnakumar and S.~Nannapaneni, ``Forecasting bike sharing demand using quantum bayesian network,'' \emph{Expert Systems with Applications}, vol. 221, p. 119749, 2023.

\bibitem{carrascal2023bayesian}
G.~Carrascal, G.~Botella, A.~del Barrio, and D.~Kremer, ``A bayesian-network-based quantum procedure for failure risk analysis,'' \emph{EPJ Quantum Technology}, 2023.

\bibitem{Vicente}
V.~P. Soloviev, C.~Bielza, and P.~Larrañaga, ``Quantum approximate optimization algorithm for bayesian network structure learning,'' \emph{Quantum Inf Process}, 2023.

\bibitem{N.Innan}
N.~Innan \emph{et~al.}, ``Quantum state tomography using quantum machine learning,'' \emph{Quantum Machine Intelligence}, vol.~6, p.~28, 2024.

\bibitem{Marco}
M.~Scutari, C.~Vitolo, and A.~Tucker, ``Learning bayesian networks from big data with greedy search: computational complexity and efficient implementation,'' \emph{Statistics and Computing}, vol.~29, pp. 1095--1108, 2019.

\bibitem{data}
\BIBentryALTinterwordspacing
``Oil spill classification.'' [Online]. Available: \url{https://www.kaggle.com/datasets/sudhanshu2198/oil-spill-detection}
\BIBentrySTDinterwordspacing

\bibitem{javadi2024quantum}
A.~Javadi-Abhari \emph{et~al.}, ``Quantum computing with qiskit,'' \emph{arXiv preprint arXiv:2405.08810}, 2024.

\end{thebibliography}
\end{document}